# Gravitational radiation in dynamical noncommutative spaces

S. A. Alavi, M. Amiri Nasab

*Department of Physics, Hakim Sabzevari University,*
*P. O. Box 397, sabzevar, Iran*
*s.alavi@hsu.ac.ir , alaviag@gmail.com*

**Abstract**

We investigate gravitational radiation in dynamical noncommutative spaces. By including corrections to the gravitational potential due to dynamical noncommutativity, we calculate the power in gravitational radiation and use observational data to place an upper bound on the noncommutativity parameter. We also study quantum interference induced by gravitational potential in the usual and dynamical noncommutative spaces, and compare the resulting phase difference in these cases with that in commutative space.

## 1. Introduction

Gravity is one of the most extensively studied phenomena in physics. This research focus on gravity during the last century has led to a deeper understanding of its nature and helped us to increase our knowledge in Planck scale and find the missing link between general relativity and quantum theory. Gravitational waves (GW) have been extensively studied since their indirect discovery in the orbital decay of binary pulsar PSRB 1913+16. The effect of GW on matter is measured by the relative optical phase shift between the light paths in two perpendicular arm cavities. The GWs have been directly detected in the interference pattern created by a laser interferometer in LIGO experiment.

Noncommutative geometry is a well-known framework to implement quantum gravity effects in a physical system. The interesting point is that the upper bound on the noncommutative length scale appears to be comparable to the length difference typically induced by the GW. Therefore, noncommutativity of space may be a possible source of noise in GW detectors. Hence, the future GW detection experiments provide a good opportunity to probe the structure of the spacetime and find signatures of a possible noncommutativity of space.

The idea of noncommutative spaces was first suggested by Heisenberg as a possible way to remove the divergences in quantum field theory. In 1999, it was shown that [1,2] the noncommutative coordinates appear naturally in string theory, which resulted in extensive study of the subject since then. In recent years, there have been significant attempts to formulate physical theories in noncommutative spaces, ranging from condensed matter to high energy physics to gravity and cosmology [3-22]. The noncommutative coordinates satisfy the following commutation relation:

$$[x_\mu, x_\nu] = i\theta_{\mu\nu}, \tag{1.1}$$

where $\theta_{\mu\nu}$ is an anti-symmetric tensor. The simplest case corresponds to the situation where $\theta_{\mu\nu}$ is constant, which we call non-dynamical noncommutative space (NDNCS or $\theta$-space). In general, $\theta_{\mu\nu}$ can be a function of the coordinates [23]. A generalization to noncommutative spaces in two dimensions has been recently proposed [24], called dynamical noncommutative spaces (DNCS), in which the commutation relations are position dependent:

$$\begin{aligned}
&[Y, P_y] = i\hbar(1 + \tau Y^2) &, &\quad [X, Y] = i\theta(1 + \tau Y^2) \\
&[X, P_x] = i\hbar(1 + \tau Y^2) &, &\quad [Y, P_x] = 0 \\
&[X, P_y] = 2i\tau Y(\theta P_y + \hbar X) &, &\quad [P_x, P_y] = 0
\end{aligned} \tag{1.2}$$

It is easy to see that in the limit $\tau \to 0$, the non-dynamical noncommutative relations are recovered:

$$\begin{aligned}
&[x_\circ, y_\circ] = i\theta &\quad [x_\circ, p_{x_\circ}] = i\hbar &\quad [x_\circ, p_{y_\circ}] = 0 \\
&[y_\circ, p_{y_\circ}] = i\hbar &\quad [y_\circ, p_{x_\circ}] = 0 &\quad [P_{x_\circ}, p_{y_\circ}] = 0
\end{aligned} \tag{1.3}$$

The coordinate $X$ and the momentum $P_y$ in Eq. (1.2) are not Hermitian. So, in general, the Hamiltonian of the system is a non-Hermitian operator. However, one can find a similarity transformation, i.e., a Dyson map $\eta \mathcal{O} \eta^{-1} = o = o^\dagger$, which transforms a non-Hermitian system into a Hermitian one. It is shown that [24] the corresponding Dyson map is $\eta = (1 + \tau Y^2)^{-\frac{1}{2}}$. The new Hermitian variables $x$, $y$, $p_x$, $p_y$ satisfy commutation relations in (1.2), and can be expressed in terms of non-dynamical variables ($\theta$-space) as follows:

$$\begin{aligned}
x &= \eta X \eta^{-1} = (1 + \tau y_\circ^2)^{-\frac{1}{2}}(1 + \tau y_\circ^2)x_\circ(1 + \tau y_\circ^2)^{\frac{1}{2}} = (1 + \tau y_\circ^2)^{\frac{1}{2}}x_\circ(1 + \tau y_\circ^2)^{\frac{1}{2}} \\
y &= \eta Y \eta^{-1} = (1 + \tau y_\circ^2)^{-\frac{1}{2}}y_\circ(1 + \tau y_\circ^2)^{\frac{1}{2}} = y_\circ \\
p_x &= \eta P_x \eta^{-1} = (1 + \tau y_\circ^2)^{-\frac{1}{2}}p_{x_\circ}(1 + \tau y_\circ^2)^{\frac{1}{2}} = p_{x_\circ} \\
p_y &= \eta P_y \eta^{-1} = (1 + \tau y_\circ^2)^{-\frac{1}{2}}(1 + \tau y_\circ^2)p_{y_\circ}(1 + \tau y_\circ^2)^{\frac{1}{2}} = (1 + \tau y_\circ^2)^{\frac{1}{2}}p_{y_\circ}(1 + \tau y_\circ^2)^{\frac{1}{2}}
\end{aligned} \tag{1.4}$$

Using Bopp-shift, one can relate the $\theta$-variables to the conventional commutative space variables [25]:

$$x_{i_\circ} = x_{i_s} - \frac{\theta_{ij}}{\hbar}p_{j_s} \quad , \quad p_{i_\circ} = p_{i_s} \quad , \quad i,j = x,y, \tag{1.5}$$

where $\theta_{ij} = \varepsilon_{ijk}\theta_k$.

It is worth noting that the indices "∘" and "s" denote variables in the noncommutative $\theta$-space and the commutative space respectively. An interesting point to note is that for simultaneous measurement of $X$, $Y$ in the dynamical noncommutative space we have [24]:

$$\Delta X_{min} = \theta\sqrt{\tau}\sqrt{1 + \tau\langle Y\rangle_\rho^2} \tag{1.6}$$

This leads to a minimal length for X, while there is no non-vanishing minimal length for Y. This important result implies that in the two-dimensional dynamical noncommutative space objects are naturally of string type.

## 2. Gravitational potential in a dynamical noncommutative space

We now study a two-body system in DNCS. The gravitational potential is given by:

$$V = -\frac{Gm_a m_b}{r} = -\frac{Gm_a m_b}{[x^2+y^2]^{\frac{1}{2}}} \tag{2.1}$$

where $x$ and $y$ satisfy the relations in (1.4). In terms of $\theta$-variables, the potential reads:

$$V(x_\circ, y_\circ, p_{x_\circ}, p_{y_\circ}) = -Gm_a m_b\left[(1+\tau y_\circ^2)^{\frac{1}{2}} x_\circ (1+\tau y_\circ^2) x_\circ (1+\tau y_\circ^2)^{\frac{1}{2}} + y_\circ^2\right]^{-\frac{1}{2}}$$

$$= -Gm_a m_b[(1+\tau y_\circ^2)^2 x_\circ^2 + 4i\theta\tau y_\circ(1+\tau y_\circ^2)x_\circ - \tau\theta^2 - 2\tau^2\theta^2 y_\circ^2 + y_\circ^2]^{-\frac{1}{2}}, \tag{2.2}$$

while it has the following form in terms of standard variables:

$$V(x_s, y_s, p_{x_s}, p_{y_s}) = -G\frac{m_a m_b}{(x_s^2+y_s^2)^{\frac{1}{2}}} + G\frac{m_a m_b}{2(x_s^2+y_s^2)^{\frac{3}{2}}}\left[2\tau x_s^2 y_s^2 - \frac{2\theta}{\hbar}L_{z_s}\right] = V^s + V^{\tau,\theta} \tag{2.3}$$

This is the same as corrections to the Coulomb potential derived in [26]. Since the noncommutativity parameters $\theta$ and $\tau$ should be very small, one can always treat the noncommutative effect as a small perturbation to the commutative counterpart. Here we only consider the first order terms in $\theta$ and $\tau$.

## 3. Two-body problem in DNCS

The Hamiltonian of the system is given by:

$$H = \frac{(p_{xa}^2+p_{ya}^2)}{2m_a} + \frac{(p_{xb}^2+p_{yb}^2)}{2m_b} - \frac{k}{\sqrt{(x_a-x_b)^2+(y_a-y_b)^2}} \tag{3.1}$$

Where $k = Gm_am_b$. In terms of a new set of coordinates defined as follows:

$$x = x_a - x_b \quad , \quad x_{cm} = \frac{m_a x_a + m_b x_b}{m_a + m_b} \tag{3.2}$$

$$y = y_a - y_b \quad , \quad y_{cm} = \frac{m_a y_a + m_b y_b}{m_a + m_b}$$

The Hamiltonian is:

$$H = \frac{p_c^2}{2(m_a+m_b)} + \frac{p^2}{2\mu} - \frac{k}{\sqrt{|x|^2+|y|^2}} \quad , \quad R = \sqrt{|x|^2 + |y|^2} \tag{3.3}$$

where $\mu$ and $p_c$ are the reduced mass and momentum of the center of mass, respectively. Assuming that $p_c = 0$, in terms of the standard coordinates, we have:

$$H_{\theta,\tau} = \frac{p_s^2}{2\mu} - \frac{k}{R} + \frac{k}{2R^3}\left[2\tau x_s^2 y_s^2 - \frac{2\theta}{\hbar} L_z\right] + \frac{\tau}{\mu} y_s^2 p_{y_s}^2 - 2i\frac{\tau\hbar}{\mu} y_s p_{y_s} - \frac{\tau\hbar^2}{2\mu}. \tag{3.4}$$

The equation of motion for a circular motion follows:

$$\frac{\partial H_{\theta,\tau}}{\partial R} = \frac{k}{R^2} + k\tau \cos^2 w_0 t \sin^2 w_0 t + \frac{k\theta\mu w_0 \cos\alpha}{\hbar}\frac{1}{R^2}$$

$$+ 4\tau\mu w_0^2 \sin^2 w_0 t \cos^2 w_0 t R^3 - 4i\hbar\tau w_0 \sin w_0 t \cos w_0 t R - \mu w_{\tau,\theta}^2 R = 0 \tag{3.5}$$

where $\alpha$ is the angle between $\vec{\theta}$ and $\vec{L}$. $L = |\vec{L}| = \mu R^2 w_0$ is the angular momentum of the system in commutative space and $w_0^2 = G\frac{m_a+m_b}{R^3} = \frac{k}{\mu R^3}$. From (3.5), the angular velocity of the system in the case of DNCS is found to be:

$$w_{\tau,\theta} = w_0\left(1 + \frac{1}{2}\frac{\theta\mu w_0 \cos\alpha}{\hbar} + \frac{5}{2}\tau R^2 \cos^2 w_0 t \sin^2 w_0 t - \frac{2i\hbar}{\mu}\sin w_0 t \cos w_0 t\right) \tag{3.6}$$

and its time average is given by:

$$\overline{w}_{\tau,\theta} = w_0 + \frac{1}{2}\theta\mu w_0^2 \cos\alpha + \frac{5}{16}\tau R^2 w_0 \quad , \quad \hbar = 1. \tag{3.7}$$

## 4. Period decay of a two-body system

For a two-body system consisting of masses $m_a$ and $m_b$ in the $x^3 = 0$ plane, and in the center-of-mass frame, one has:

$$\begin{aligned} x_a^1 &= -x_b^1 = \mu \Re \cos w_{\tau,\theta} t \\ x_a^2 &= -x_b^2 = \mu \Re \sin w_{\tau,\theta} t \\ x_a^3 &= x_b^3 = 0 \end{aligned} \qquad (4.1)$$

where $\Re$ is given by:

$$\Re^2 = R_s^2 - 2\vec{L}.\vec{\theta} + \frac{1}{4}\tau R_s^4 \quad , \quad R_s^2 = x_s^2 + y_s^2 \qquad (4.2)$$

The rate of energy loss or the total luminosity is $p = \frac{G}{45}\langle \dddot{D}^{ij}\dddot{D}^{ij}\rangle$ [27], where the quadrupole moments are given by:

$$D^{11} = \mu\left(R^2 - 2\vec{L}.\vec{\theta} + \frac{1}{4}\tau R^4\right)(3\cos^2 w_{\tau,\theta}t - 1) \qquad (4.3)$$

$$D^{12} = D^{21} = 3\mu\left(R^2 - 2\vec{L}.\vec{\theta} + \frac{1}{4}\tau R^4\right)\sin w_{\tau,\theta}t \cos w_{\tau,\theta}t$$

$$D^{22} = \mu\left(R^2 - 2\vec{L}.\vec{\theta} + \frac{1}{4}\tau R^4\right)(3\sin^2 w_{\tau,\theta}t - 1)$$

$$D^{23} = -\mu\left(R^2 - 2\vec{L}.\vec{\theta} + \frac{1}{4}\tau R^4\right)$$

The mean total luminosity of the system $P_{\tau,\theta} = P_{\tau,\theta}^a + P_{\tau,\theta}^b$ is then as follows:

$$P_{\tau,\theta} = -\langle\frac{dE_{\tau,\theta}}{dt}\rangle = \frac{32}{5c^5}G\mu^2 R^4 w_{\tau,\theta}^6 \left(1 - 4\mu w_0 \theta \cos\alpha + \frac{1}{2}\tau R^2\right)$$

$$= \frac{32}{5c^5}G\mu^2 R^4 w_0^6 \left(1 - \theta\mu w_0 \cos\alpha + \frac{19}{8}\tau R^2\right). \qquad (4.4)$$

Defining $w_0^2 = \frac{k}{\mu R^3}$, the energy of the system is:

$$E_{\tau,\theta} = \frac{1}{2}\mu w_0^2 R^2 - \frac{k}{R} + \frac{1}{4}k\tau R - \frac{k}{R}\theta\mu w_0 \cos\alpha + \frac{\tau}{2\mu} = -\frac{1}{2}\frac{k}{R} + \frac{1}{4}k\tau R - \sqrt{\frac{k^3\mu}{R^5}}\theta\cos\alpha + \frac{\tau}{2\mu} \qquad (4.5)$$

and the rate of energy loss is given by (note that $\frac{dE_{\tau,\theta}}{dt} = \frac{dE_{\tau,\theta}}{dR}\frac{dR}{dt}$):

$$\frac{-dE_{\tau,\theta}}{dt} = \frac{-1}{2}\frac{k}{R^2}\left(1 + 5\theta\mu w_0 \cos\alpha + \frac{1}{2}\tau R^2\right)\frac{dR}{dt} \qquad (4.6)$$

After using (4.4) and (4.6), we have:

$$\frac{dR}{dt} = \frac{-64}{5c^5}\left(\frac{G}{R}\right)^3 (m_a m_b)(m_a + m_b)\frac{\left[1-\theta\mu w_0 \cos\alpha + \frac{19}{8}\tau R^2\right]}{\left[1+5\theta\mu w_0 \cos\alpha + \frac{1}{2}\tau R^2\right]}.\tag{4.7}$$

By setting $\theta$ and $\tau$ to zero, we recover the results in the commutative space [27].

Using (4.7) and $\eta = \dot{T} = 3\pi\sqrt{\frac{\mu R}{k}}\dot{R}$, we can derive the rate of the period decay as follows:

$$\eta_{\tau,\theta} = \frac{-192}{5c^5}\pi\left(\frac{2\pi}{T}G\right)^{\frac{5}{3}}\frac{m_a m_b}{(m_a m_b)^{\frac{1}{3}}}\frac{\left(1-\theta\mu w_0 \cos\alpha + \frac{19}{8}\tau R^2\right)}{\left(1+5\theta\mu w_0 \cos\alpha + \frac{1}{2}\tau R^2\right)} = \eta_0 + \Delta\eta \tag{4.8}$$

where $\eta_0 = \frac{-192}{5c^5}\pi\left(\frac{2\pi}{T}G\right)^{\frac{5}{3}}\frac{m_a m_b}{(m_a m_b)^{\frac{1}{3}}}$, and $\Delta\eta = -6\theta\mu w_0 \eta_0 \cos\alpha + \frac{15}{8}\tau\eta_0 R^2$ is the correction to the period decay rate due to noncommutativity of space.

## 5. The two-body system 1913+16 PSR

The mass of the pulsar 1913+16PSR and its companion are:

$$m_p = 1.44 \, M_\odot$$
$$m_c = 1.38 \, M_\odot \tag{5.1}$$

where $M_\odot$ is the solar mass and $e = 0.61$ is the eccentricity of the orbit. For a non-circular orbit $(e \neq 0)$, the rate of the orbital decay $\eta_0$ includes the factor $f(0.61) = 11.85$ [28]. The values of theoretical and observational orbital decay rate and the period of the system are [28]:

$$\eta_0 = -2.42 \times 10^{-12} \, s/s$$
$$\eta_0^{obs} = -2.40 \times 10^{-12} \, s/s \tag{5.2}$$
$$T = 27895.56 \, s$$

For an elliptic orbit, there is a minimum $R_{min} = 1.1 \, R_\odot$ and a maximum $R_{max} = 4.8 \, R_\odot$ for the radius, where $R_\odot = 6.955 \times 10^5 km$ is the solar radius. After using the constraint:

$$|\Delta\eta| < |\eta_0 - \eta_0^{obs}| \tag{5.3}$$

and the numerical values of relevant quantities, we obtain the following upper bound on the dynamical noncommutative parameter $\tau$:

$$\sqrt{\tau} < 10^{-16} \, ev \, . \tag{5.4}$$

It is worth mentioning that one can also find an upper bound on the non-dynamical parameter $\theta$ along the same line [29].

It is interesting to note that $\sqrt{\theta}$ has dimension of length ($L$), while $\sqrt{\tau}$ has the dimension of energy ( or $L^{-1}$, see Eq. (1.2) ) and it is a measure of the impact of the dynamical noncommutativity of space on the energy of the system. The bound on $\sqrt{\tau}$ in (5.4) is consistent with the accuracy in the energy measurement $10^{-12}$ eV [30], and this is reasonable because there is no evidence for space noncommutativity yet.

The interaction of GW with matter is very weak. For instance the cross-section for the interaction of GW with the hydrogen atom is of the order of $10^{-66}$ cm$^2$[31], which is extremely small. GW are also very hard to detect. As a gravitational wave passes by, objects would change their length, but this effect is incredibly small. So we have never before been able to measure them and this is why gravitational waves detected 100 years after Einstein's prediction. Therefore it is not surprising if the corrections on gravitational interaction due to space noncommutativity is very small.

## 6. Gravity-induced quantum interference

First, we briefly review the problem in the standard case of commutative space. Consider a nearly mono-energetic beam of particles (e.g., neutrons) split into two parts that are brought back together as shown in Fig.(1). Let us consider two paths A→ B→ D and A→ C→ D that lie in a horizontal plane. Because the absolute zero of the gravitational potential is of no significance, one can set V=0 in this plane. Therefore, gravity does not induce any phase difference between the two paths in this case. However, if we rotate this plane about AD by an angle $\delta$, there will be a gravity-induced phase difference between ABD and ACD paths [32]. In general, the phase difference between two paths is given by:

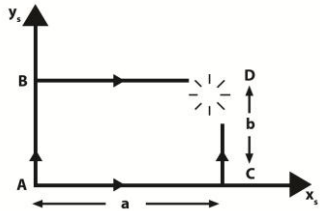

Fig (1). Experiment to detect gravity-induced quantum interference with a horizontal plane.

$$\Delta\varphi = \varphi_2 - \varphi_1 = \frac{1}{\hbar}[V_2(t) - V_1(t)]T \qquad (6.1)$$

where T is the time spent for the wave packets to go from A to D.

Now, let us consider the same paths and study the effect of dynamical and non-dynamical noncommutativity of space on the phase difference induced by gravity between the amplitudes of the two wave packets arriving at D. The coefficients proportional to θ and $\tau$ in Eq. (2.3) show the effects of non-dynamical and dynamical noncommutativity on the gravitational potential respectively. In the $\tau$-space, one can divide the path ACD into two segments AC and CD. By choosing the point A as the origin of the coordinates, and noting that $y_s = 0$ for the segment AC, we see that the contribution of this segment is zero. For the segment CD, since $x_s = a$ but $y_s$ varies, we can integrate over $y$ in Eq. (2.3) to find the following contribution to the gravitational phase for the path ACD:

$$k\tau a^2 \int_0^b \frac{y^2}{(a^2+y^2)^{3/2}} dy = k\tau a^2 \left\{ \log[b + \sqrt{a^2+b^2}] - \frac{a}{\sqrt{a^2+b^2}} - \frac{\log[a^2]}{2} \right\} \quad (6.2)$$

Similarly, for the path ABD we find:

$$k\tau b^2 \int_0^a \frac{x^2}{(x^2+b^2)^{3/2}} dx = k\tau b^2 \left\{ \log[a + \sqrt{a^2+b^2}] - \frac{b}{\sqrt{a^2+b^2}} - \frac{\log[b^2]}{2} \right\} \quad (6.3)$$

Thus the phase difference induced by gravity between the amplitudes for the two wave packets arriving at D in the $\tau$-space is given by:

$$(6.4)$$

$$V_\tau = \left\{ k\tau \left[ b^2 \log(a + \sqrt{a^2+b^2}) - a^2 \log(b + \sqrt{a^2+b^2}) \right] + \frac{ab}{\sqrt{a^2+b^2}}(a-b) - \frac{1}{2}(b^2 \log b^2 - a^2 \log a^2) \right\}$$

It is seen that, contrary to the commutative space, there is a phase difference between the two paths in the $\tau$-space even for a horizontal plane. It is worth mentioning that this contribution vanishes when a=b.

We now consider the same problem in the $\theta$-space. For the two line segments AC and CD, we have:

$$L_z|_{AC} = xp_y - yp_x = 0 \quad (6.5)$$
$$L_z|_{CD} = ap - 0 = ap \quad (6.6)$$

Similarly, for the segments AB and BD, we find:

$$L_z|_{AB} = 0$$
$$L_z|_{BD} = -bp \quad (6.7)$$

where $|p_x| = |p_y| = p$. This results in

$$\Delta L_z = ap - (-bp) = (a+b)p \quad (6.8)$$

For nonrelativistic particles $E = \frac{p^2}{2m}$, and hence $\Delta L_z = (a + b)\sqrt{2mE}$. The interesting point here is that while the gravity-induced phase difference between two horizontal paths in the $\tau$-space is energy independent, it depends on the particle energy in the $\theta$-space. Hence, by varying the beam energy one may verify whether the noncommutativity of space is dynamical or non-dynamical.

Next, we consider the same problem for a horizontal circle of radius R in the noncommutative $\theta$-space shown in Fig. (2).

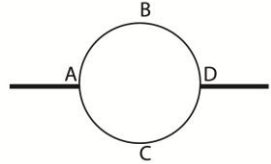

Fig (2). Experiment to detect gravity-induced quantum interference with a horizontal circle.

According to Eq. (2.3), the correction to the gravitational potential is proportional to $L_z$. For the two paths in Fig. (2), $L_z$ has the same magnitude but with opposite signs. If $L_z = m\hbar$ for the path ACD, it will be $-m\hbar$ for ABD resulting in $\Delta L_z = 2m\hbar$. We then have:

$$\Delta\varphi \propto \frac{m\hbar\theta}{R^3} T \tag{6.9}$$

In the $\tau$-space $x_s$ and $y_s$ vary for both of the paths ACD and ABD, which makes integration necessary:

$$\frac{k\tau}{R^3} \int x_s^2 dx_s \int y_s^2 dy_s \tag{6.10}$$

It is more convenient to perform the integral over $\theta$ using the well-known relations $x_s = R\cos\theta$ and $y_s = R\sin\theta$. For the first integral, one has:

$$\int y_s^2 dy_s = \int_\pi^0 R^2 \sin^2\theta (R\cos\theta d\theta) = 0 \tag{6.11}$$

Thus, the phase difference induced by gravity between the two paths ABD and ACD on a horizontal circle is zero in the $\tau$-space. We summarize the results in Table 1 below.

|  | Horizontal plane | Horizontal circle |
|---|---|---|
| Commutative space | 0 | 0 |
| $\theta$-space | $\neq 0$<br>But depends on the beam energy | $\neq 0$ |
| $\tau$-space | $\neq 0$<br>No dependence on the beam energy | 0 |

Table 1. The summary of gravity induced phase difference.

## 7. Discussion

In this section, we would like to discuss some interesting issues:

### 7.1. Comparison to other methods used to study the effects of noncommutativity on cosmological observations

In this paper, we have used the observational limits from the binary systems to constrain the space time noncommutativity parameter. As shown in [33], the noncommutativity of space can also leave its imprints on the CMB anisotropy. This is because the noncommutativity of space becomes important at extremely small length scales, which could be relevant during an epoch of inflation in the very early universe.

This issue has also been studied in [34]. It is therefore interesting to make a comparison between their method and ours and check their consistency. It is well understood that the various quantum mechanical relations can be obtained from the corresponding classical relations by replacing the Poisson brackets in the latter by commutation relations in the former:

$$[\,,\,]_{\text{classical}} = \frac{[\,,\,]}{i\hbar} \tag{7.1}$$

The classical Poisson brackets and the quantum mechanical commutators satisfy similar algebraic properties. The best example is the cornerstone relation of quantum mechanics:

$$[x_i, p_j] = i\hbar \delta_{ij} \tag{7.2}$$

which corresponds to the classical Poisson bracket $[x_i, p_j]_{\text{classical}} = \delta_{ij}$. Another example is the angular momentum whose components satisfy $[L_i, L_j]_{\text{classical}} = \epsilon_{ijk} L_k$ in classical mechanics, which in quantum mechanics turns into:

$$[L_i, L_j] = i\hbar \epsilon_{ijk} L_k \tag{7.3}$$

This approach is very powerful because it can be generalized to situations where observables have no classical analog. For example, the spin components has nothing to do with space coordinates and momentum components in classical mechanics, but in quantum mechanics its components satisfy the following commutation relations:

$$[S_i, S_j] = i\hbar \epsilon_{ijk} S_k \tag{7.4}$$

which can be derived using the properties of rotations (see [32] for more discussions).

The important point to note is that there are no contradictions between these relations. When we are dealing with a quantum mechanical problem, we just decide which of the relations in (7.2), (7.3) or (7.4) is more suitable for the problem at hand.

In [33,34], as explicitly stated in [33], the authors followed a formalism similar to [35] (see also [36]) in order to include the noncommuativity and imposed the following deformed commutation relation on the conjugate momenta:

$$[\pi_{ij}(\tau,\vec{K}),\pi_{kl}(\tau,\vec{K})] = \theta_{ijkl}\delta^{(3)}(\vec{K}-\vec{K}) \quad (7.5)$$

where $\theta_{ijkl}$ is an anti-symmetric tensor defined as $\theta_{ijkl} = \theta_{ik}\delta_{jl} + \theta_{il}\delta_{jk} + \theta_{jk}\delta_{il} + \theta_{jl}\delta_{ik}$ where $\theta_{jk} = -\epsilon_{0jkl}\theta_l$, $\theta_l$ is a constant 3-vector. The zero in the Levi-Civita tensor shows that the authors considered only the noncommutativity between space-space coordinates. For convenience, the authors took $\theta_1 = \theta_2 = 0$ and $\theta_3 \neq 0$ in [33]. It is necessary to mention that in [33,34] the authors used $\alpha_{ijkl}$ instead of $\theta_{ijkl}$, which was used in [35]. In [35], in order to impose the noncommutativity on the system, the authors promote the metric $h_{ij}$ and the conjugate momenta $\pi_{ij}$ into operators whose commutation relations are obtained by considering the simplest deformation of the classical Poisson brackets:

$[\pi_{ij}, \pi_{kl}]_{classical} = 0$

$[h_{ij}, h_{kl}]_{classical} = 0$

$[h_{ij}(\vec{K}), \pi_{kl}(\vec{K})]_{classical} = \frac{1}{2}(\delta_{ik}h_{jl} + h_{jk}h_{il})\delta(\vec{K}-\vec{K}) \quad (7.6)$

as follows:

$[\pi_{ij}, \pi_{kl}] = \theta_{ijkl}$

$[h_{ij}, h_{kl}] = 0$

$[h_{ij}(\vec{K}), \pi_{kl}(\vec{K})] = \frac{1}{2}(\delta_{ik}h_{jl} + h_{jk}h_{il})\delta(\vec{K}-\vec{K}) \quad (7.7)$

Hence, the method used in [33-36] to go from the commutative to the noncommutative case is just what we do in quantum mechanics as mentioned earlier. Therefore, depending on the nature of the problem under consideration, one can start from the fundamental relation in Eq. (1.1) (our method), or use the commutation relations (7.7), without any contradiction between the two methods. Here we would like to point out that [15,21,37] have used the same method as ours to study gravitational waves in noncommutative spaces.

### 7.2. Comparison to other works based on noncommutative spectral geometry

We would also like to make a comment on the paper by Nelson et.al. [38] and its relation to our work. The paper is based on noncommutative spectral geometry (NCSG), which itself is based on spectral action introduced by Connes and Chamseddine [39]. Spectral geometry concerns relationships between the geometric structure of manifolds and the spectra of canonically defined differential operators. In NCSG, a noncommutative geometric space is encoded by a spectral triple (A, H, D) where the algebra A is the algebra of functions that interact with the inverse line element D, by acting in the same Hilbert space H, where D is an unbounded self-adjoint operator [40].

It is explicitly stated in [41] that the NCSG approach is compatible with the noncommutative approach based upon $[x_i, x_j] = i\theta_{ij}$. The reason being that in the literature a noncommutative space is often of Moyal type, involving noncommutative tori or Moyal planes and the Euclidean version of Moyal noncommutative field theory is compatible with the spectral triples formulation of noncommutative geometry. In the language of spectral geometry let $A := C^\infty(T_\theta)$ be a smooth noncommutative torus associated with a non-zero skew-symmetric deformation $\theta$. $C^\infty(T_\theta)$ is generated by some unitary elements $u_i$, which are subject to the relations :

$$u_i u_j = e^{i\theta_{ij}} u_j u_i \tag{7.8}$$

or:

$$u_i u_j = e^{\frac{i}{2}\theta_{ij}} u_{i+j} \tag{7.9}$$

As shown in several papers, see e.g, [2,42] these relations lead to $[x_i, x_j] = i\theta_{ij}$. It is also interesting to mention that by starting from commutation relation $[x_i, x_j] = i\theta_{ij}$, defining the basic plane wave [43] as:

$$u_i = \exp(ix_i) \tag{7.10}$$

and using the Baker-Campbell-Hausdorff formula, we can show that these operators generate the algebra in (7.8) or (7.9).

In summary, the spectral action and NCSG are based on spectral triples and noncommutative tori and their subalgebras generate the basic commutation relation in (1.1), which in turn generates the basic algebra of a torus in NCSG in (7.8) and (7.9). Thus, the method used in this paper and the NCSG method used in [38] are consistent. It is worth mentioning that the parameter β in [38] that is constrained by using astrophysical limits has a similar role to $\frac{1}{\theta}$ in our work. In the limit β → ∞, the NCSG results are reduced to those of the (ordinary) general relativity.

### 7.3. Gravity –induced quantum interference pattern

The third issue we would like to comment on is what physical observables might be measured in the lab related to gravity-induced quantum interference in a noncommutative space discussed in Section 6.

In a horizontal plane (circle) in a commutative space there is no phase difference between the two beams arriving at point D (Δφ = 0), and therefore there is no interference pattern. However, in the presence of the noncommutativity of space, the phase difference between the paths ABD and ACD is not zero, which results in an observable interference pattern. Another interesting point is that, for a horizontal plane, for non-dynamical noncommutativity the interference pattern does not depend on the energy of particles, while for dynamical noncommutativity the pattern changes with varying the energy.

The situation in our experiment is somehow similar to the LIGO experiment because in both cases the physical observable is the interference pattern. In LIGO, the quantity measured by the detector is the net phase shift:

$$\Delta\varphi = \delta\varphi_1 - \delta\varphi_2 \qquad (7.11)$$

which is the difference between the individual phase shifts experienced by the light in each of the two arms. In the absence of a gravitational wave signal, $\Delta\varphi = 0$ [44]. When a gravitational wave passes through the interferometer, the length of the two arms change and the laser beams travel different distances. In our case we use neutrons instead of photons to form interference pattern.

In the last century interferometry was a valuable tool for studying special relativistic effects. In the 21$^{st}$ century we expect a new revolution where interferometry techniques can probe the interplay between the quantum world and gravity.

**7.4. Variation of the angle in gravity –induced quantum interference experiment**

Finally, we would like to emphasize that the plane (circle) considered in Section 6 is horizontal, and hence there is no "variation of the angle" in our study.

**8. Conclusion**

In string theory, the point-like particles are replaced by one-dimensional objects (strings). String theory is a promising candidate for quantum gravity, thereby unifying all forces in nature. As mentioned in the text, objects in the dynamical noncommutative space that we studied here are naturally string like, which implies that DNCS has a deeper relation to string theory than NDNCS. We have also shown that some operators in DNCS are non-Hermitian. The relations between DNCS and the theory of non-Hermitian operators from one hand, and DNCS and string theory from the other hand, may lead to new insights and achievements in all the three fields.

In this paper, we studied gravitational radiation in DNCS. We obtained corrections to the angular velocity and radiated power of a two-body system due to dynamical noncommutativity. We then calculated the period decay of the system and used observational limits to place an upper bound on the DNCS parameter. The corrections due to non-dynamical noncommutativity ($\theta$-corrections) on the periodic decay rate found in the literature depend on $\cos\alpha$, where $\alpha$ is the angle between $\vec{\theta}$ and $\vec{L}$, and hence vanish for $\alpha = \frac{\pi}{2}$ (unlike the DNCS case). We also studied the gravitational phase shift induced by gravity for two paths in a horizontal plane. We showed that contrary to the commutative case, there is a phase difference between the two paths in both $\theta$-space and $\tau$-space cases. Therefore high-precision gravitational interferometry techniques can be used to determine whether there exists noncommutativity of space in nature and, in which case, its dynamical or non-dynamical origin.

**Acknowledgments**

S. A. Alavi would like to thank "INFN, Sezione di Torino and Dipartimento di Fisica" for kind hospitality and support during his visit to INFN where part of this work was done. We are also very grateful to Dr. Estiri and Prof. Rouzbeh Allahverdi (university of New Mexico, USA) for careful reading of the manuscript and valuable comments.